# Breaking the Cycle of Recurring Failures: Applying Generative AI to Root Cause Analysis in Legacy Banking Systems

Siyuan Jin[*]
HKUST
Hong Kong, China

Zhendong Bei[*]
HSBC
Guangzhou, China

Bichao Chen[*]
HSBC
Guangzhou, China

Yong Xia[†]
HSBC
Guangzhou, China

## Abstract
Traditional banks face significant challenges in digital transformation, primarily due to legacy system constraints and fragmented ownership. Recent incidents show that such fragmentation often results in superficial incident resolutions, leaving root causes unaddressed and causing recurring failures. We introduce a novel approach to post-incident analysis, integrating knowledge-based GenAI agents with the "Five Whys" technique to examine problem descriptions and change request data. This method uncovered that approximately 70% of the incidents previously attributed to management or vendor failures were due to underlying internal code issues. We present a case study to show the impact of our method. By scanning over 5,000 projects, we identified over 400 files with a similar root cause. Overall, we leverage the knowledge-based agents to automate and elevate root cause analysis, transforming it into a more proactive process. These agents can be applied across other phases of the software development lifecycle, further improving development processes.

## 1 Introduction
Traditional banks face challenges in their digital transformation, driven by existing legacy systems and their strict regulations [2]. Legacy systems are not flexible to adapt to market changes, and risk-averse cultures further hinder innovation [9, 12]. As a result, some legacy systems generate repeated incidents, significantly impacting bank operations. For example, the Monetary Authority of Singapore (MAS) imposed a six-month pause on DBS Bank's non-essential activities due to operational failures[1]. To remain competitive in a fast-changing, customer-focused financial market, traditional banks need to adopt more flexible and integrated technology strategies [2].

In legacy systems, incident management often requires cross-team collaboration. Teams may spend much time debating who is responsible for an incident, focusing on surface-level symptoms instead of the actual root causes. While incidents eventually get fixed, the failure to address the core issues leads to recurring problems over time. This raises the key research question of our study: *How can post-incident analysis methods be improved to identify root causes and prevent similar problems from happening again?*

As systems become more complex and interconnected, finding the root cause of failures becomes harder. Traditional Root Cause Analysis (RCA) methods, like the "Five Whys" [11], often miss the multi-layered nature of these systems. These methods rely heavily on human judgment, which introduces bias and variability. In large IT operations, particularly those using legacy systems, the focus is often on addressing symptoms rather than the deeper, systemic problems. This results in incomplete fixes and recurring incidents. Also, the large volume of incident data in complex environments overwhelms traditional RCA methods, limiting their scalability. As a result, incidents are often treated in isolation, leading to reactive problem-solving and increasing technical debt.

To address this, we integrate the traditional "Five Whys" RCA with GenAI. We use a knowledge graph to capture knowledge from the entire software development lifecycle, providing a knowledge base for GenAI agents. By leveraging GenAI, we analyze both problem description data and evidence sources (i.e., actions taken), enabling a thorough exploration of not only what teams have documented but also the actions they have implemented.

We find that over 70% of issues previously attributed to external factors—such as management—were caused by internal code deficiencies or gaps in automation. We demonstrate how our root cause can avoid recurring failures via a case example and how similar code patterns exist in other legacy systems, which facilitates proactive code quality improvement. Our knowledge-based agents provide implications for other processes as well. They can aid in tasks such as code generation, and documentation validation. These models led to a 45% reduction in major incidents over a year, a 45.5% reduction in change failure rate, and a 46.3% decrease in lead time to deployment within a global finance company.

## 2 Background
### 2.1 Fragmented Ownership of Incidents
Fragmented ownership in legacy systems impedes effective root cause identification during incident management. Figure 1 illustrates a typical flawed workflow where incidents are logged and categorized without thoroughly investigating their underlying causes.

---
[*]Three authors contributed equally to this research.
[†]Corresponding Author.
[1]https://www.mas.gov.sg/news/media-releases/2023/mas-imposes-six-month-pause-on-dbs

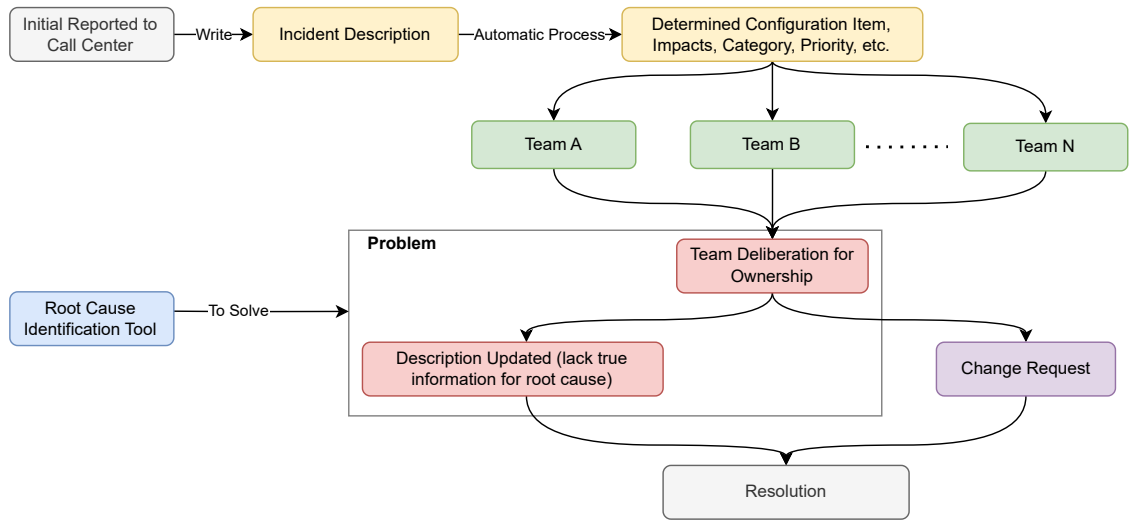

Figure 1: Flawed Incident Management Process

Table 1: Repeated Symptom Issues Example

| Issue Category | Incident IDs | Recurring Issue Description |
| --- | --- | --- |
| Long-running SQL Queries | INC10, INC11 | Inefficient query optimization, leading to repeated performance degradation and resource exhaustion. |
| Database Deadlocks | INC1, INC2 | Ineffective deadlock detection, resulting in unresolved locks and system delays. |
| Storage Capacity Exhaustion | INC5, INC6, INC7 | Inadequate capacity monitoring, leading to space exhaustion and service disruptions. |
| Backup Failures | INC8, INC9 | Systemic failures in backup processes, affecting disaster recovery. |

Responsibility is dispersed across multiple teams, each focusing on isolated components of the system. This siloed approach prevents a holistic understanding of the problem, leading to superficial fixes rather than addressing systemic issues, even when formal RCA methods are applied. Teams often implement temporary solutions to restore service, but no single team is accountable for resolving the deeper issues. As a result, technical debt increases, and operational costs rise due to repeated failures.

Table 1 presents key recurring issues from our dataset. These problems highlight the inefficiency of current incident management processes, where fragmented ownership limits the scope of analysis to immediate symptoms (i.e., the focus is on short-term fixes rather than long-term solutions). This escalates operational risks and perpetuates inefficiencies, making future system failures more likely. To address these, Generative AI (GenAI) offer a promising solution. By integrating insights across the fragmented software development components, GenAI can provide a more comprehensive understanding of the system's behavior, enabling scalable and more accurate incident resolution.

### 2.2 GenAI in Software Development

Figure 2 shows how knowledge is generated during the Software Development Life Cycle (SDLC) and utilized by agents. Each SDLC phase produces artifacts that contain valuable entity information and relationships. We developed a procedure to automatically extract and map this data, constructing a knowledge graph that supplies accurate, real-time information to analytical agents. This knowledge graph enhances downstream processes by providing a structured, integrated view of the system. As an illustration, we apply this approach to root cause analysis, highlighting its effectiveness in improving analysis precision and response time.

The phases of the SDLC are interdependent, with decisions in one phase influencing outcomes in subsequent phases. For instance, inadequate planning can result in unrealistic timelines and resource constraints, which in turn compress development time, limit testing, and lead to incomplete requirements gathering. Problems in early phases, such as planning or requirements, propagate through the lifecycle, increasing both the complexity and cost of remediation.

GenAI-based agents, which can sense their environment, make decisions, and take action, have shown strong performance across various domains [1]. In the SDLC, GenAI enhances each phase by automating tasks, improving decision quality, and facilitating knowledge transfer between phases [4].

In the requirements gathering phase, GenAI can reduce ambiguity in user stories, improving clarity and alignment between stakeholders [8]. By analyzing historical data, GenAI also automates project scoping and estimation, enhancing the accuracy of timelines, resource allocation, and risk assessment. These refined requirements enable early identification of potential design and testing issues, minimizing rework.

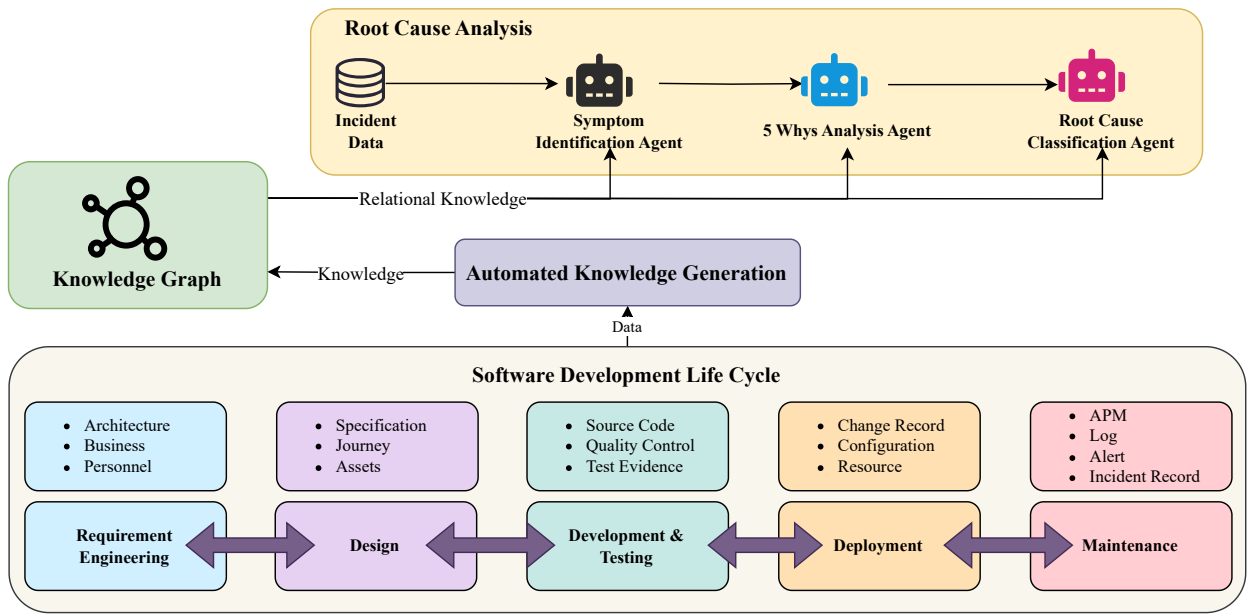

Figure 2: Knowledge-Enabled Agents in Software Development Life Cycle (SDLC)

Table 2: Comparison of Root Cause Analysis (RCA) Methods in IT

| Method | Description | Strengths | Limitations |
| --- | --- | --- | --- |
| **Five Whys [11]** | Repeatedly asks "Why?" to trace root causes. | Simple, scalable | Prone to oversimplification, human bias. |
| **FMEA [5]** | Identifies potential failure modes and assesses their impacts. | Systematic, data-driven | Requires domain expertise, resource-intensive. |
| **Fault Tree Analysis [10]** | Constructs a hierarchical model of failure causes. | Comprehensive, visualizes dependencies | Complex, time-consuming, requires expert judgment. |
| **Pareto Analysis [7]** | Prioritizes causes based on the 80/20 principle. | Focuses on key drivers | May overlook less obvious or hidden root causes. |

In the design phase, GenAI can provide automated design suggestions [14] such as architectural patterns or user interface designs, ensuring that iterations are faster and that best practices are followed. These improvements carry over into the development phase, where GenAI can generate code, automate testing, and detect bugs early in the process [6, 14]. By speeding up development and ensuring consistency, GenAI reduces testing cycles and the likelihood of defects, leading to a more streamlined development process.

In the deployment phase, GenAI automates the deployment process and continuously monitors system performance, which helps minimize errors and ensures system stability. Real-time anomaly detection during deployment further reduces the risk of failures. Data collected during deployment is then fed back into planning and design, helping to identify improvement areas in future projects.

During maintenance, GenAI leverages predictive analytics to identify potential failures in advance, reducing downtime and proactively addressing issues [15]. These insights ensure system availability is maintained and further reduce the need for reactive maintenance. Furthermore, lessons learned during maintenance, such as recurring system issues, are used to guide improvements in earlier phases, such as planning and design, leading to better risk management and more resilient systems in future projects.

## 2.3 Post-Incident Analysis and Root Cause Identification

Post-incident analysis is essential for keeping IT systems reliable, especially in critical sectors like banking, where disruptions can cause serious financial and reputational damage [13]. A key part of this process is RCA, which goes beyond fixing symptoms to find and address the underlying reasons for system failures [3].

These weaknesses are clear in environments where incidents keep happening due to unresolved systemic problems. For example, in large banking IT systems, issue resolution often focuses on immediate fixes like restarting services or patching software, rather than addressing deeper root causes. This leads to repeated incidents, higher operational risk, and reduced system resilience.

Table 2 compares the common RCA techniques in IT, such as the "Five Whys". These methods trace incidents to their root causes, identifying technical or procedural issues that need fixing to prevent future problems. However, these methods have limitations in modern, large-scale systems. They depend on human expertise, which can introduce bias and limit their scalability. Traditional methods also struggle to handle the complex interactions between technical and organizational factors common in IT systems [13]. Conversely, GenAI, by combining structured and unstructured data (such as incident logs, system metrics, and user feedback), offers a holistic approach to RCA [13].

## 3 Methodology

Our solution leverages GenAI to automate and enhance the post-incident RCA process in IT environments. A key challenge in RCA is the absence of ground truth for root causes, which often require actionable insights at the code level—far more granular than high-level incident descriptions like "service down." LLMs serve as powerful tools in this context, drawing on existing knowledge from the whole SDLC phases to provide deeper insights into root causes. By integrating AI agents into the traditional RCA workflow—specifically augmenting the 5 Whys method, we introduce the funnel model to identify underlying causes of incidents (figure 3). The solution enables faster, more accurate, and scalable RCA by utilizing LLMs and a dynamic knowledge graph that aggregates system events, logs, and incident data.

**Knowledge Graph.** We use end-to-end automation throughout the SDLC to build a knowledge graph that gathers evidence from all phases. This graph is a rich source of information, helping AI agents make better decisions across the software process.

**Symptom Analysis Agent** can collect incident data from IT service management (ITSM) systems. This agent identifies key symptoms and possible problem areas by querying the knowledge graph. It establishes a baseline for the incident, allowing the 5 Whys analysis to focus on the most relevant parts.

**Five Whys Analysis Agent** automates the iterative questioning process to identify the root cause of incidents. It systematically analyzes incident data, probing deeper with each subsequent question. Real-time data from the knowledge graph serves as evidence, supporting the investigation. We integrate the GPT-4o model with a knowledge graph for evidence retrieval, enhancing traditional analytical methods by automating workflows to pinpoint underlying issues. GPT-4o processes complex problem descriptions, guiding the analysis, while the evidence retrieval system extracts and integrates relevant data from the knowledge base, providing context and reinforcing the analytical process.

The agent's architecture leverages the React pattern, ensuring modularity and flexibility in managing state and handling effects, which optimizes the integration of diverse data sources. This design allows for consistent consolidation of datasets and facilitates the retrieval of historical data to enhance analytical depth. The integration of AI with structured evidence retrieval significantly improves both the rigor and precision of root cause investigations.

**Root Cause Classification Agent** validates the root cause identified by the Five Whys Analysis Agent by comparing it against historical patterns and classifying it into relevant categories. The agent first detects recurring patterns in historical incident data where similar problems have frequently occurred. It then utilizes advanced language models to categorize the root cause of each incident within these patterns. This approach helps uncover systemic issues, enabling more efficient resolution in future operations. By leveraging this process, the agent not only confirms the results of the Five Whys Analysis but also plays a critical role in strategic problem remediation efforts. In 95% of cases, the findings were validated by post-incident reviews, confirming its high level of accuracy in pinpointing actionable root causes.

## 4 Results

Figure 4 demonstrates the application of this process in practice. The case study outlines the step-by-step use of the AI-enhanced "Five Whys" approach. Starting with a high-level issue, such as an unreachable service, the system iterates through successive "Why?" questions to uncover the root cause. Each step is informed by real-time data from the knowledge graph, ensuring decisions are based on objective evidence. In this example, the root cause—initially attributed to service delays—was traced to a lack of automation. The issue was ultimately resolved through a targeted change request, ensuring a permanent fix and minimizing the risk of recurrence. This case study illustrates how the AI-enhanced "Five Whys" not only identifies root causes with greater accuracy but also facilitates durable solutions, reducing future incidents.

We compare the "Five Whys" method, enhanced by generative AI (GenAI), with traditional root cause analysis (RCA) techniques. As shown in Figure 5, the AI-enhanced "Five Whys" method significantly shifts problem attribution. Our analysis found that over 70% of issues previously attributed to external factors—such as management errors or vendor failures—were actually due to internal code deficiencies or automation gaps. Traditional approaches often misattribute issues to external sources, whereas the AI-supported "Five Whys" identifies a higher percentage of internal causes, enabling more precise and effective resolutions. This shift highlights the value of data-driven, AI-enhanced RCA in complex IT environments.

## 5 Discussion

This section outlines the impact of our model with one case study. As shown in **Table 3**, before implementation, one request system experienced significant performance issues, such as response times exceeding 1,200ms and frequent service failures. After implementation, these issues were resolved, with all requests consistently completed in under 800ms and no failures reported. Scanning 5,535 projects and identifying recurring defects in 415 files, the AI has showed its scalability and effectiveness across diverse environments.

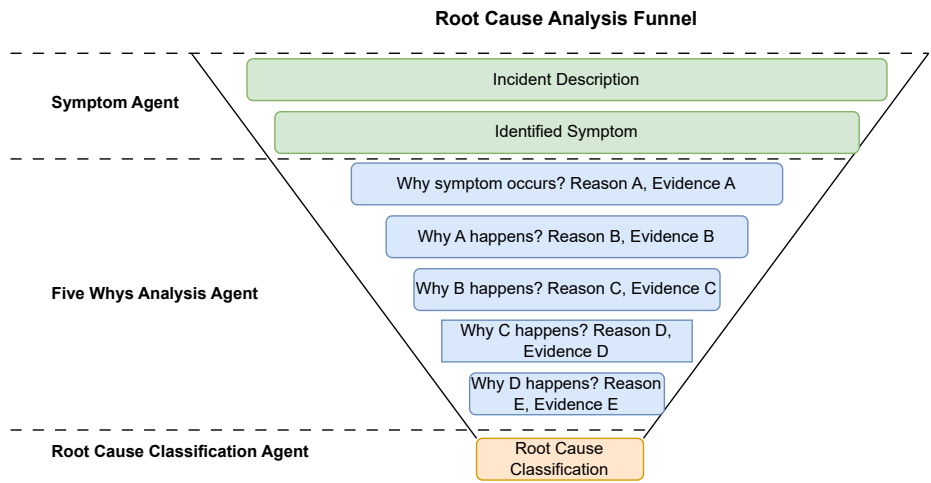

Figure 3: Our Proposed Funnel Model

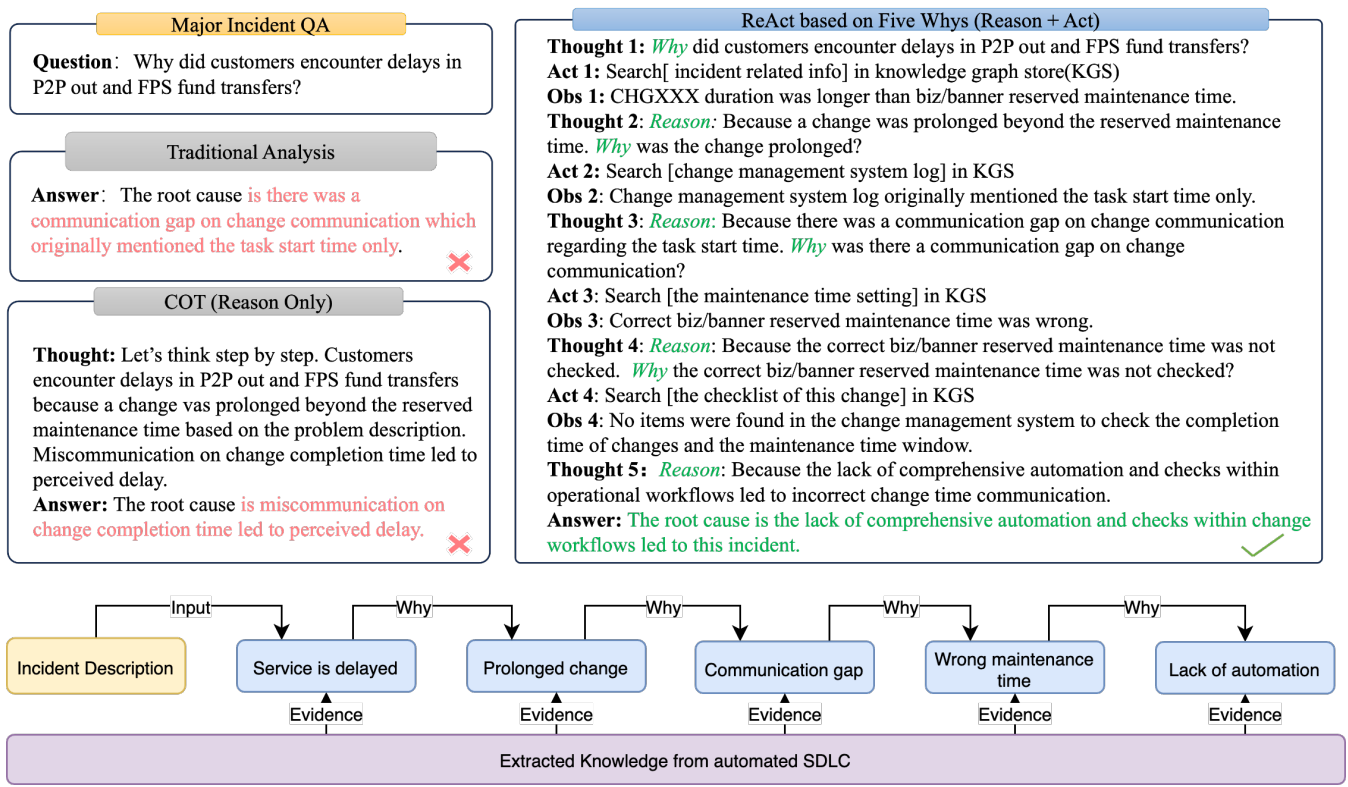

Figure 4: Case Study

This capacity to handle large volumes of data with precision ensures its applicability in enterprise-scale operations, reducing the need for manual oversight in defect management.

Overall, the success of this GenAI-driven system highlights the broader potential of using AI to automate defect management. By integrating AI into the SDLC, organizations can ensure continuous code improvement while reducing manual effort and minimizing human error. The system's ability to scale, adapt, and enhance risk management sets a strong precedent for future AI-driven software development practices. As AI technology evolves, its role in predictive maintenance and real-time security will likely grow, further improving software reliability and resilience.

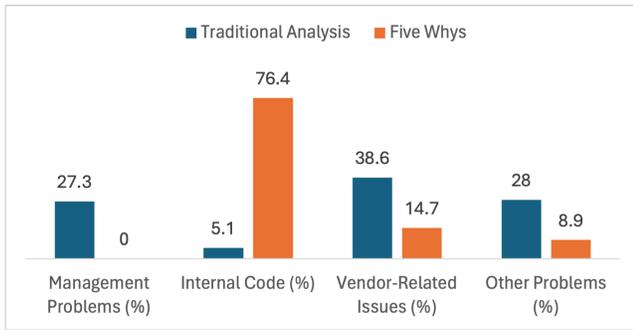

Figure 5: Comparison of Classification Results

Table 3: Case Summary

| Aspect | Details |
|---|---|
| **Total Projects** | 5,535 |
| **Same-Defect Projects** | 226 |
| **Same-Defect Files** | 415 |
| **Pre-Fix Performance** | - Response times > 800ms, peaking at 1,200ms.<br>- Frequent failures. |
| **Post-Fix Performance** | - All requests < 800ms.<br>- No failures. |

## 6 Conclusion

The interconnected nature of software management means that disruptions in one phase can have cascading effects throughout the entire lifecycle. Our study has demonstrated that integrating knowledge-based agents into this process can significantly improve efficiency and reliability. The models we introduced led to measurable improvements, including a 45% reduction in major incidents, a 45.5% drop in change failure rate, and a 46.3% decrease in lead time to deployment within a global finance company.

Beyond these quantitative gains, our approach offers broader benefits for operational efficiency and software management. By automating the detection and resolution of code defects, the system reduces the manual workload on developers, allowing them to focus on higher-value tasks. This automation accelerates the defect resolution process and enhances the consistency and reliability of patches, leading to higher overall code quality. As a result, the system not only boosts immediate performance but also supports sustainable software development practices, reinforcing long-term resilience and reducing technical debt.

While our model has demonstrated significant success, future research should explore its scalability across different industries and system architectures. Additionally, addressing potential limitations, such as AI biases or edge cases where the model may be less effective, will be critical for ensuring its broader applicability. Nevertheless, the integration of Generative AI into root cause analysis offers a promising direction for improving software management, ensuring that organizations can adapt to increasingly complex and dynamic IT environments.